\documentclass{iopart}
\usepackage{graphicx}
\usepackage{iopams}
\usepackage{subfigure}
\begin{document}

%title
\title[Search for a stochastic GW signal in the MLDC]
  {Search for a stochastic gravitational-wave signal in the second 
  round of the Mock LISA Data Challenges}

\author{E~L~Robinson$^{1}$, J D~Romano$^{2,3}$ and A~Vecchio$^{1}$}

\address{${}^1$ School of Physics and Astronomy, 
University of Birmingham, Edgbaston, Birmingham B15 2TT, UK}
\address{${}^2$ School of Physics and Astronomy,
Cardiff University, Cardiff CF24 3AA, Wales, UK}
\address{${}^3$ Department of Physics and Astronomy 
and Center for Gravitational-Wave Astronomy,
University of Texas at Brownsville, Brownsville, TX 78520, USA}

\eads{
\mailto{elr@star.sr.bham.ac.uk},
\mailto{joe@phys.utb.edu},
\mailto{av@star.sr.bham.ac.uk}
}

\begin{abstract}
The analysis method currently proposed to search for isotropic 
stochastic radiation of primordial or astrophysical origin with 
the Laser Interferometer Space Antenna (LISA) 
relies on the combined use of two LISA channels, one of which 
is insensitive to gravitational waves, such as the symmetrised Sagnac. 
For this method to work, it is essential to know how the instrumental 
noise power in the two channels are related to one another; 
however, no quantitative estimates of this key information are 
available to date. 
The purpose of our study is to assess the performance of the 
symmetrised Sagnac method for different levels of prior information 
regarding the instrumental noise. 
We develop a general approach in the framework of Bayesian 
inference and an end-to-end analysis algorithm based on Markov 
Chain Monte Carlo methods to compute the posterior probability
density functions of the relevant model parameters. 
We apply this method to data released as part of the second round of 
the Mock LISA Data Challenges. 
For the selected (and somewhat idealised) cases considered here, 
we find that a prior uncertainty of a factor $\approx 2$ in the 
ratio between the power of the instrumental noise contributions 
in the two channels allows for the detection of isotropic stochastic 
radiation. 
More importantly, we provide a framework for more realistic studies 
of LISA's performance and development of analysis techniques in the 
context of searches for stochastic signals.
\end{abstract}

%\maketitle

\section{Introduction}
\label{s:introduction}

The Laser Interferometer Space Antenna (LISA) 
\cite{lisa} will observe many different gravitational-wave signals, 
including stochastic gravitational radiation from a variety of sources. 
The gravitational-wave 
foreground produced by galactic white dwarf binaries is a 
guaranteed (anisotropic) signal \cite{Hills-et-al:1990, BenderHills:1997, HillsBender:2000, Nelemans-et-al:2001, Edlund-et-al:2005}, 
but one also expects LISA to see other types of stochastic 
gravitational radiation, possibly including (isotropic) 
backgrounds from the very early universe~\cite{Maggiore:2000, Hogan:2006, DamourVilenkin:2005, Siemens-et-al:2007}  
and foregrounds from 
populations of extra-galactic sources~\cite{FarmerPhinney:2003, BarackCutler:2004-emriconf}. 
For anisotropic foregrounds, the stochastic signal 
will be modulated, with a period of one year, by LISA's 
orbital motion around the Sun.
Hence search techniques that take advantage of this modulation 
can be used to distinguish the astrophysical signal from 
instrumental noise
\cite{UngarelliVecchio:2001, Cornish:2001, Seto:2004, 
SetoCooray:2004, KudohTaruya:2005, TaruyaKudoh:2005, Edlund-et-al:2005}.
For isotropic backgrounds, an alternative search method is 
needed, which will work in the absence of any (statistical) 
time-variation of the signal relative to the noise.

One approach that has been proposed to search for isotropic 
gravitational radiation is to use a particular combination of 
the LISA output, called the \emph{symmetrised Sagnac} channel
\cite{TintoEtAl:2000}, as a real-time noise monitor for LISA.
This is possible at low-frequencies (i.e., less than a few mHz), 
since the response of the symmetrised Sagnac channel 
to gravitational waves is highly suppressed 
for these frequencies.
One can then compare the power in the symmetrised Sagnac channel
to that in another observable that {\em is} sensitive to 
gravitational waves, and
then `subtract' the two to estimate the power in the
gravitational-wave background
\cite{TintoEtAl:2000,HoganBender:2001}.
For this subtraction method to work, one needs to know how the 
instrumental noise power in the two channels are related to one 
another.
Otherwise, one might falsely assign too much (little) power 
to the
gravitational-wave component if one underestimates 
(overestimates) the 
instrumental noise power in the gravitational-wave
channel.
Obviously, one cannot unambiguously observe isotropic stochastic 
radiation without any prior knowledge about the instrumental noise. 

In this paper, we extend the analyses of
\cite{TintoEtAl:2000,HoganBender:2001} by recasting the 
symmetrised Sagnac method in the framework of Bayesian inference;
we then use it to search for stochastic gravitational-wave 
signals in data taken from round 2 of the Mock LISA 
Data Challenge (MLDC) \cite{mldc, mldclisasymp, mldcgwdaw2, mldcamaldi2}.
(These are mock data sets released by the MLDC Taskforce, which 
contain simulated data from simplified models of LISA and 
restricted numbers of astrophysical sources.)
The purpose of our study is to assess the performance 
of the symmetrised Sagnac method for different levels of 
prior information regarding the instrumental noise.
% in the context of the MLDCs.
Using Markov Chain Monte Carlo (MCMC) methods to calculate the
posterior probability density functions (PDFs) of the relevant
parameters, we explicitly show how the uncertainty in 
our estimate of the gravitational-wave signal power 
depends on our prior knowledge of the relationship between the 
instrumental noise power in the symmetrised Sagnac and 
gravitational-wave channels. In particular, we find that a prior 
uncertainty 
of a factor $\approx 2$ in the ratio between the power of
the instrumental noise contributions in the two channels allows 
for the detection of 
isotropic stochastic radiation. For the specific case
considered here (and driven by the MLDC data sets released in Challenge 2), 
in which the amplitude of the gravitational-wave signal 
dominates the instrumental noise,
a more accurate knowledge of this ratio, say within a factor
of $\approx 10\%$, 
provides only a marginal improvement for signal detection. 
This particular result depends 
of course on the signal-to-noise ratio, and we do not try,
in this paper,
to answer conclusively this question of detectability as a 
function of signal-to-noise ratio and prior information. 
Nonetheless, we provide a conceptually transparent framework that 
allows us to quantitatively investigate this problem; the formalism enables 
more realistic studies of LISA's performance 
and associated requirements in the
context of searches for stochastic signals that could yield some of the most 
exiting discoveries of the mission. 
Some of these issues are currently
under investigation and will be discussed in detail in a forthcoming paper \cite{RobinsonEtAl:2008}.

The rest of the paper is organised as follows:
In Sec.~\ref{s:method}, we present details of the analysis 
method, and its application to LISA. 
In Sec.~\ref{s:results}, we give the results of our analysis 
when applied to the MLDC data sets.
Finally, in Sec.~\ref{s:conclusion} we summarise our 
findings and discuss possible extensions for future investigations. 

%%%%%%%%%%%%%%%%%%%%%%%%%%%%%%%%%%%%%%%%%%%%%%%%%%%%%%%%%%%%%%%%%%

\section{Analysis method}
\label{s:method}

\subsection{LISA time-delay interferometry combinations $A$, $E$, and $T$}

LISA will consist of three spacecraft in a (nearly) equilateral configuration 
of side $5\times 10^6$~km, trailing the Earth by about 20 degrees. 
The distances between the spacecraft will be modulated by incident 
gravitational waves at 
the level of picometres.
The modulation will be sensed by monitoring the frequency
(or, equivalently, phase) of laser beams exchanged between the spacecraft, 
and comparing this to locally-generated reference laser signals.

%For frequencies below $\sim\!\!10\ {\rm mHz}$, 
The six raw Doppler measurements 
can be combined in various ways using the principle of 
Time-Delay Interferometry (TDI) \cite{TDI}. 
We use the first-generation TDI combinations, in which it is
assumed that LISA is rigid and symmetric, and that 
laser frequency noise cancels completely. This is consistent with 
the conventions adopted in the context of the MLDCs
\cite{mldclisasymp, mldcgwdaw1, mldcgwdaw2, mldcamaldi2, mldcgwdaw12},
but the analysis presented here can be generalised to second-generation
TDI. In particular, we work with the $A$, $E$ and $T$ combinations,
which are independent and noise-orthogonal \cite{Prince-et-al:2002}. 
For frequencies smaller than the inverse of the light
travel-time down a LISA arm 
($1/16.6\ {\rm s}\simeq 10\ {\rm mHz}$), 
the $A$ and $E$ combinations are equivalent to 
two unequal-arm Michelson 
interferometers, with independent noise, 
rotated at 45 degrees to each other, and thus are
sensitive to the two orthogonal 
polarisations of gravitational waves.
For these low frequencies,
the response of the $T$ channel to
gravitational-waves is highly-suppressed (similar to the
response of the standard symmetrised Sagnac combination $\zeta$), 
and hence can be used as a real-time noise monitor, 
as discussed above.
For simplicity, in what follows, we will restrict 
attention to the $A$ and $T$ channels.
The analysis can easily be extended to also include $E$,
which will improve our ability to detect gravitational waves
by reducing the uncertainty in our estimates by
a factor of $\sqrt{2}$ \cite{RobinsonEtAl:2008}.

%%%%%%%%%%%%%%%%%%%%%%%%%%%%%%%%%%%%%%%%%%%%%%%%%%%%%%%%%%
\subsection{Bayesian inference}
\label{s:details}

For low frequencies, the output of the $A$ and $T$ channels 
in the frequency domain have the form
\begin{equation}
\tilde{A} = \tilde{n}_A + \tilde{h}\,,
\qquad
\tilde{T} = \tilde{n}_T\,,
\end{equation}
where $\tilde{n}_{A,T}$ denote
the instrumental noise in the two channels, and 
$\tilde{h}$ denotes the stochastic gravitational-wave signal.
(Here $\widetilde{\ }$ denotes the discrete Fourier transform of the
time domain data, so that $\tilde A$, $\tilde T$, etc.\ are 
dimensionless quantities.) Notice that we
assume that in the frequency window of interest
the gravitational-wave contribution
is perfectly suppressed in the $T$ channel.
We also assume that the instrumental noises and stochastic signal
are zero-mean Gaussian random variables, with variances
\begin{equation}
\langle|\tilde{n}_A|^2\rangle=\sigma_A^2\,,
\qquad
\langle|\tilde{n}_T|^2\rangle=\sigma_T^2\,,
\qquad
\langle|\tilde{h}|^2\rangle=\sigma_h^2\,,
\end{equation}
and that the noises in the two channels are related by 
\begin{equation}
\sigma_A^2 = a\sigma_T^2\,,
\label{e:def-a}
\end{equation}
where $a$ is some multiplicative factor, which we may not 
know in advance.
(The variances $\sigma_A^2$, $\sigma_T^2$, $\sigma_h^2$, 
and multiplicative factor $a$ all vary as functions of 
frequency; however, in the analysis below
we consider one frequency bin at a time, 
and therefore do not explicitly show the dependence
of these variables on frequency.)
Since the noise in the $\tilde A$ and $\tilde T$ 
channels are 
uncorrelated with one another and with the gravitational-wave 
signal, the covariance matrix is
\begin{equation}
C = \left(
\begin{array}{cc}
a\sigma_T^2 + \sigma_h^2 & 0           \\ 
0                        & \sigma_T^2
\end{array}
\right).
\label{e:cov}
\end{equation}
Note that the problem consists, in general, of {\em three} 
unknown variables, 
$\vec\theta\equiv(\sigma_h^2, \sigma_T^2, a)$.
Measurements of the two channels 
$\vec{d}\equiv(\tilde{A},\tilde{T})$ give us only {\em two} 
independent constraints.

Application of Bayes' theorem
allows us to construct the joint posterior probability 
density function (PDF) of the unknown parameters
\begin{equation}
p(\vec{\theta}|\vec{d}) = 
\frac{p(\vec{\theta})
p(\vec{d}|\vec{\theta})}
{p(\vec{d})}\,,
\end{equation}
where $p(\vec{\theta})$ 
is the joint prior PDF of the parameters,
$p(\vec{d}|\vec{\theta})$ 
is the likelihood function,
and $p(\vec{d})$, the so-called evidence or marginal likelihood, can be 
regarded simply as a normalisation constant throughout.
The likelihood function for a single frequency bin 
is a multivariate Gaussian
\begin{equation}
%p(\tilde{A},\tilde{T}|\sigma_h^2,a,\sigma_T^2,{\cal I}) = 
p(\vec d|\vec\theta) = 
\frac{1}{2\pi} \frac{1}{\sqrt{\det{C}}} 
\exp{\left[ -\frac{1}{2} \sum_{i,j=1}^2 d_i^{*} [C^{-1}]_{ij} d_j \right]}\,,
\end{equation}
where $C$ is the noise covariance matrix given by (\ref{e:cov}).
We can construct a likelihood function for multiple 
time segments by simply multiplying the likelihoods of the 
individual segments.

The posterior PDF for any given parameter 
(or subset of parameters), 
say $\theta_1$, can be obtained by {\em marginalising} the 
joint posterior over the remaining parameters:
\begin{equation}
p(\theta_1 | \vec{d}) = 
\int d\theta_2 \int d\theta_3\>p(\vec{\theta} | \vec{d})\,.
\label{e:posterior-marg-gen}
\end{equation}
Markov Chain Monte Carlo (MCMC) methods can be used to explore the 
posterior PDFs, and to find the marginal posterior PDF on each parameter. 
From the 
marginal posteriors one can then compute the posterior mean on each 
parameter:
\begin{equation}
\bar{\theta}_i = \int_{-\infty}^{\infty} 
d \theta_i\>
\theta_i p(\theta_i|\vec{d})\,,
\end{equation}
and the $95\%$ probability intervals 
$[\theta_{i,{\rm low}},\theta_{i,{\rm high}}]$, 
where the limits define the {\em smallest} interval 
containing $95\%$ of the total probability:
\begin{equation}
0.95 = \int_{\theta_{i,{\rm low}}}^{\theta_{i,{\rm high}}} 
d \theta_i\>
p(\theta_i|\vec{d})\,.
\end{equation}

The prior PDFs contain the information 
we have about the parameters before making the observations. 
For our analysis we assume {\em no prior knowledge} of the variances 
$\sigma_h^2$ and $\sigma_T^2$, and choose flat priors
\begin{equation}
p(\sigma_h^2) \propto \left\{
\begin{array}{ll}
1&\mathrm{for\,\,\,}0<\sigma_h^2<\sigma_{h,\mathrm{max}}^2\\
0&\mathrm{otherwise},
\end{array}\right.
\end{equation}
and similarly for $\sigma_T^2$. 
We also used flat priors for $a$, but in order to investigate 
how our prior knowledge of the multiplicative
factor $a$ affected the posterior 
PDFs, the upper and lower limits of the priors were varied,
to represent different states of prior knowledge of the 
relative instrumental noise levels in the two channels.

%%%%%%%%%%%%%%%%%%%%%%%%%%%%%%%%%%%%%%%%%%%%%%%%%%%%%%%%%%%%
\section{Mock LISA Data Challenge results}
\label{s:results}

\subsection{MLDC data sets}
\label{s:datasets}

The analysis method detailed above has been applied to the 
second round of the MLDC~\cite{mldcgwdaw2}. 
The data sets supplied by the MLDC Taskforce are the $X$, $Y$,
and $Z$ ``unequal-arm Michelson'' combinations of the six 
optical readouts.
The $A$, $E$ and $T$ combinations which we use can be constructed 
in the 
frequency domain as linear combinations of $X$, $Y$ and $Z$. 

The second round of the MLDC consists of various data sets containing
different classes of signals \cite{mldc, mldcgwdaw2}. 
The 2.1 data set, which we were primarily interested in, 
contains the signal from $\approx 26$ million white dwarf binaries
generated according to a Galactic population model superimposed 
on Gaussian-stationary instrumental noise. 
Although the astrophysical population produces
an anisotropic stochastic signal below a few mHz, with 
some sources generating sufficiently
strong radiation to allow their identification, 
we apply our method to the training 2.1 data set
as a proof-of-principle demonstration of the symmetrised Sagnac approach. 
Since the background is anisotropic,
the level of the galactic signal will vary throughout the year as 
LISA orbits the Sun, so our method provides an estimate of the 
{\em average} power of the
gravitational-wave signal over the total observation time.

\begin{figure}
\centering\includegraphics[width=4in,angle=0]{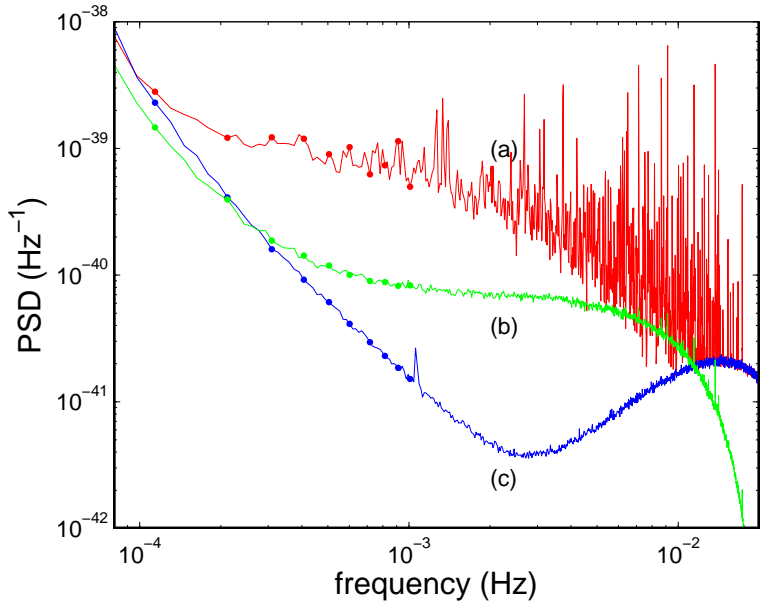}
\label{f:mldc_noise}
\caption{Power spectral densities (PSDs) 
obtained from the MLDC 2.1 and 1.1.1a data sets. 
Line (a) shows the PSD of the $A$ channel in the 2.1 data set, which
includes the instrumental noise and an injected galaxy of $\approx 26$ 
million white dwarf binaries. Line (b) shows the PSD of the
$T$ channel in the 2.1 data set. Line (c) shows the PSD
of the $A$ channel in the MLDC 1.1.1a data set, which contains just 
one injected binary, at $1.06$ mHz, which is outside of the frequency 
band of interest. 
As there are no signals in the 1.1.1a data set below $\sim 1$ mHz, 
this PSD is 
representative of the detector noise in the $A$ channel, and was 
used to determine the true value of $a$, which was needed to 
validate the results of our analysis.}
\end{figure}

One important point that we addressed was how to validate the results of our analysis. 
The first round data set 1.1.1a~\cite{mldc, mldclisasymp} contains just instrumental noise
(with the same statistical properties and spectrum as the one used for the 2.1 data sets), with one 
injected binary signal at 1.06 mHz, which is outside of the frequency 
range of our analysis. This data set could therefore be used to estimate 
the noise power spectral density (PSD) of the $A$ channel 
in the absence of a signal, and hence determine the true value of 
$a$, which was needed to validate our results. 
Figure \ref{f:mldc_noise} shows the PSDs of
the instrumental noise in both $A$ and $T$ channels, as well as the 
PSD of the noise and the galactic signal in the $A$ channel.

The analysis described in Sec.~\ref{s:method} was carried out on the first
year of the 2.1 training data set. As the noises and signal are {\em coloured}
(i.e., frequency-dependent)
Gaussian stochastic processes, the analysis was carried-out on one 
frequency bin at a time. We assumed that the signal was stationary, although
in reality it varies periodically over the year. The time series was split 
into segments of length $T_{\rm seg}=61440$s and sampling period
$\Delta t=15\ {\rm s}$, which were then Fourier 
transformed and the bin to study was selected. 

%%%%%%%%%%%%%%%%%%%%%%%%%%%%%%%%%%%%%%%%%%%%%%
\subsection{Single-bin analysis}
\label{s:single-bin}

To begin our analysis, 
we considered a single frequency bin centred at 
0.602 mHz, and chose three different priors for $a$, 
see Equation~(\ref{e:def-a}):
(i) an {\em unconstrained} prior, where $a$ ranged from 0 to
$a_{\rm max}$, which was several orders of magnitude 
greater than the true value of $a$;
(ii) a {\em weakly-constrained} prior, where $a$ ranged from 0
to twice the true value; and
(iii) a {\em strongly-constrained} prior, where $a$ was known
to within 10\% of the true  value.
These three priors on $a$ correspond to no information, 
moderate information, and rather accurate information about the
relationship between the instrumental noise levels in the 
$A$ and $T$ channels.
For this particular analysis, the `true' value of $a$ 
could be 
estimated by $P_A(f)/P_T(f)=0.43$ at 0.6~mHz, using the 1.1.1a data 
set, since the $A$ channel for the 1.1.1a data contains 
no signal in this frequency region.
This allows us to assess the performance of the 
analysis method as a function of our a~priori knowledge of 
the instrumental noise levels. 
(Of course, for a real analysis, the prior that we use for 
$a$ will be based largely on theoretical models of the 
instrument, its expected performance, and data 
from on-board monitoring channels
that provide information on different subsystems.)
The numerical values that we used for the priors are listed
in Table \ref{t:0p6mHz}.
\begin{table}
\begin{indented}
\item[]\begin{tabular}{lr @{$ < a < $} l}
\br
&\multicolumn{2}{l}{Prior on $a$ at 0.602 mHz} \\
\mr
`unconstrained' & 0  &  1000 \\
`weakly-constrained' & 0 & 0.86 \\
`strongly-constrained' & 0.39 & 0.47 \\
\br
\end{tabular}
\end{indented}
\caption{\label{t:0p6mHz}
The three different priors on $a$ in the 0.602 mHz bin.}
\end{table}

\begin{figure}
\centering
\subfigure[]{\includegraphics[width=2.5in,angle=0]{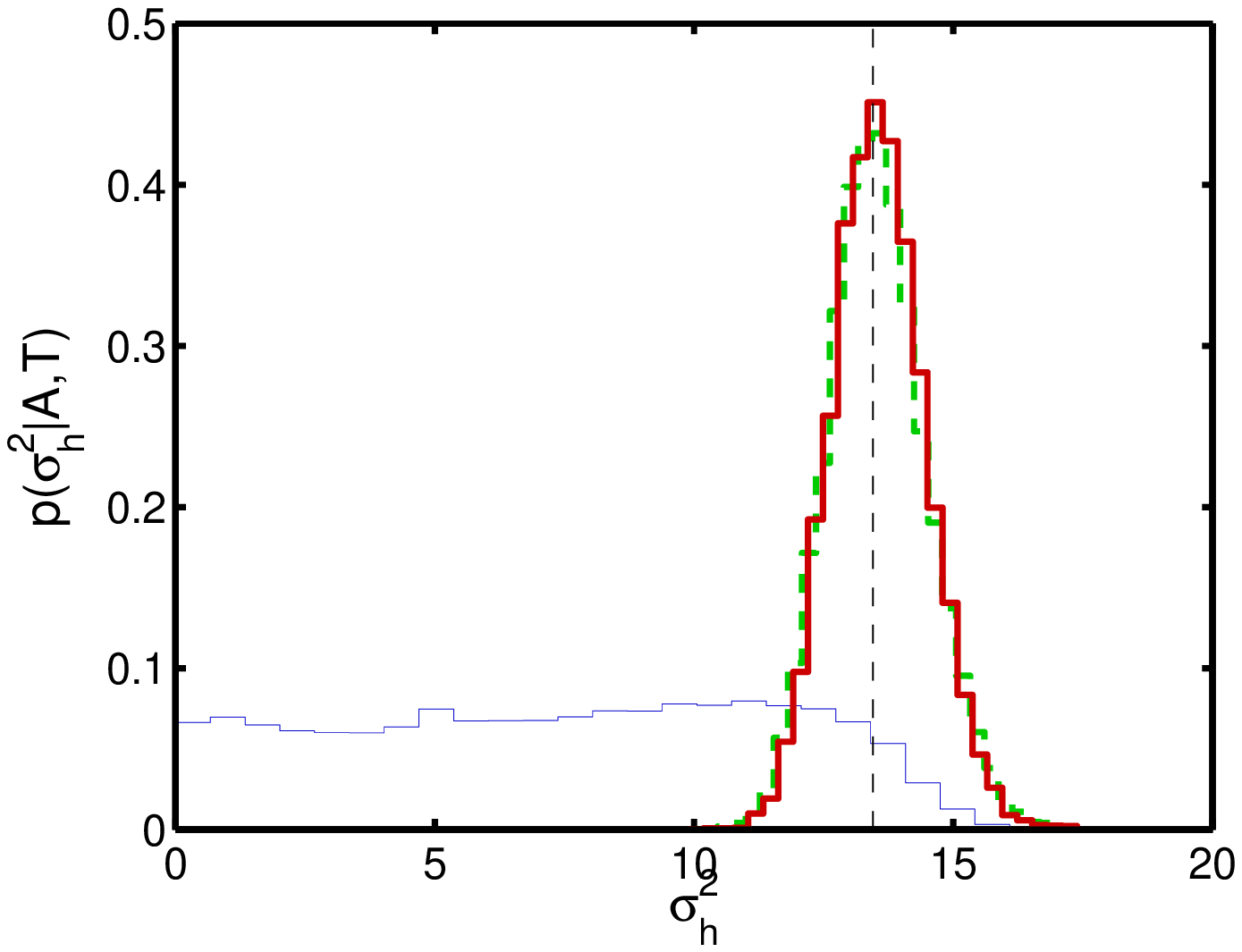}}
\subfigure[]{\includegraphics[width=2.5in,angle=0]{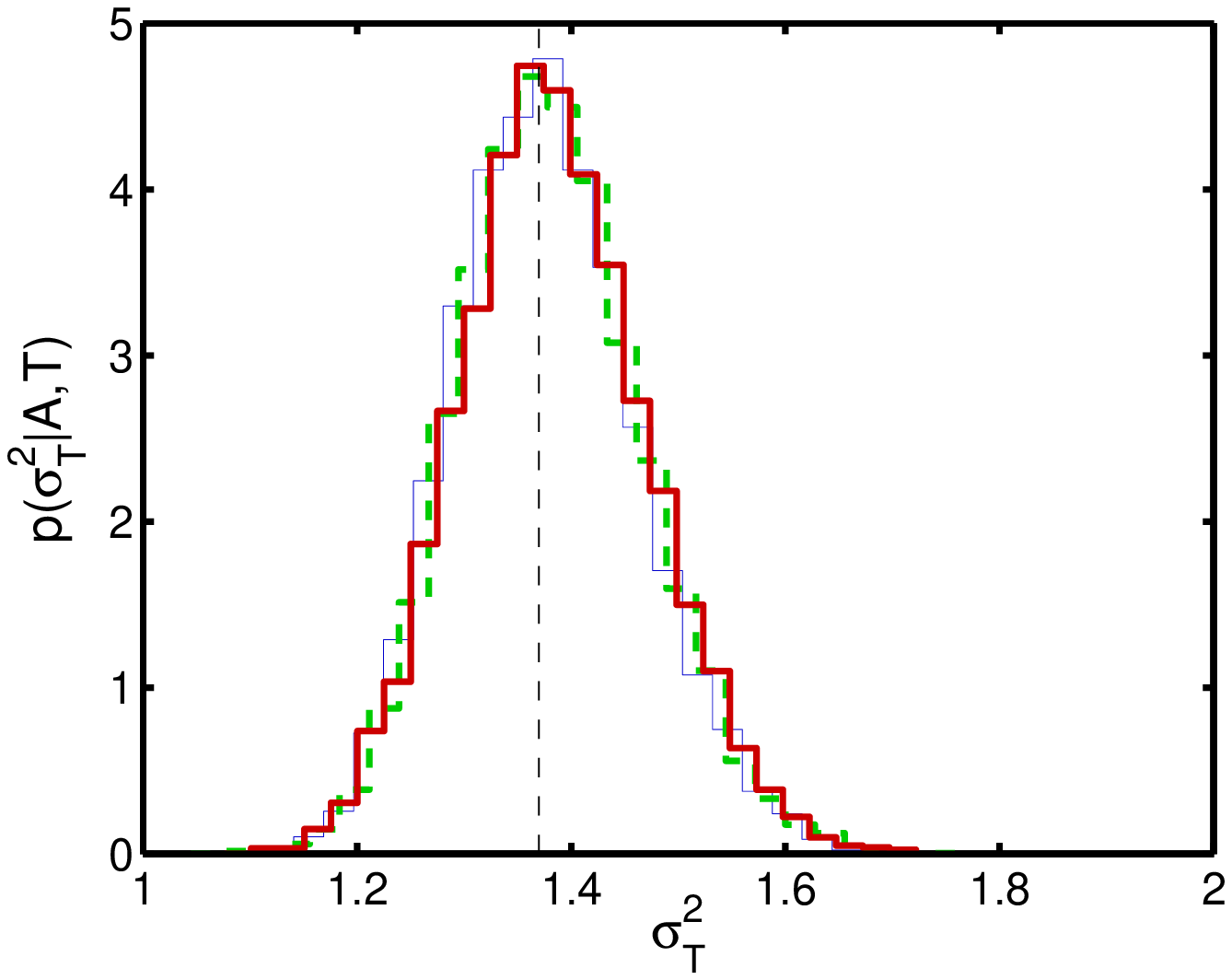}}
\caption{\label{f:post}The marginalised posterior PDFs on (a) $\sigma_h^2$ and 
(b) $\sigma_T^2$ for the three different priors on $a$, using the MLDC 2.1
data. The thin 
line shows the posterior for the unconstrained prior on $a$, the dashed
line for the weakly-constrained prior, and the thick line
for the strongly-constrained prior. The vertical dashed
lines indicate the expected values of the variances, obtained 
from the 1.1.1a and 2.1 data sets.
Notice that for 
$\sigma_T^2$, the choice of prior on $a$ makes no difference to the 
posterior.
For $\sigma_h^2$, no prior knowledge of $a$ means that we cannot 
distinguish the signal from the noise, 
while moderate knowledge (i.e., to within a factor of 2) enables us to 
easily distinguish the gravitional-wave signal and place constraints
on its value.
}
\end{figure}

Figure~\ref{f:post} shows the resultant posterior PDFs for $\sigma_h^2$ and 
$\sigma_T^2$ for the three different priors distributions on $a$. 
It can be seen that 
the PDF on $\sigma_T^2$ is not dependent on the prior on $a$. This is
to be expected, as $\sigma_T^2$ can be estimated from the $T$ channel alone, 
with no dependence on the data in the $A$ channel. The 
posteriors on $\sigma_h^2$ show that with no knowledge of $a$ it is 
impossible to determine the presence of the gravitational-wave signal. 
However, for this case,
having moderate knowledge of $a$, even to within a factor $\approx 2$, 
enables us not only to distinguish the gravitational-wave signal, 
but also to place constraints on its value. 
Using the `strongly-constrained' prior on
$a$ does not improve the result in any appreciable manner. 
This is due to the fact
that the stochastic signal is strong at this frequency, 
and in fact $\sigma_h/\sigma_{A} \approx 25$. 
The prior knowledge on $a$ required to unambiguously detect a stochastic
background clearly depends on the signal-to-noise ratio; this is an important
point and work is currently ongoing to address this 
question~\cite{RobinsonEtAl:2008}.

%%%%%%%%%%%%%%%%%%%%%%%%%%%%%%%%%%%%
\subsection{Power spectrum estimation}

In order to estimate the spectra of the gravitational-wave signal 
and the instrumental noise, and to gain some initial quantitative insight 
into the performance of the method as a function of signal-to-noise ratio, 
we studied nine other frequency bins 
distributed evenly throughout the range 
$0.1\ {\rm mHz} <f< 1.0\ {\rm mHz}$. 
As before, priors on $a$ were determined using the 1.1.1a data.
Strongly constrained priors, for each of these frequency bins, 
are given in Table \ref{t:priors}.

\begin{table}
\begin{indented}
\item[]\begin{tabular}{cr @{$ < a < $} l}
\br
Frequency&\multicolumn{2}{c}{Prior on $a$} \\
\mr
0.114 mHz & 1.33 & 1.63 \\
0.212 mHz & 1.04 & 1.27 \\
0.309 mHz & 0.75 & 0.91 \\
0.407 mHz & 0.56 & 0.68 \\
0.505 mHz & 0.46 & 0.56 \\
0.602 mHz & 0.39 & 0.47 \\
0.716 mHz & 0.29 & 0.35 \\
0.814 mHz & 0.23 & 0.29 \\
0.911 mHz & 0.20 & 0.24 \\
1.009 mHz & 0.16 & 0.20 \\
\br
\end{tabular}
\end{indented}
\caption{\label{t:priors}Strongly constrained priors on $a$ for
the different frequency bins studied.}
\end{table}

\begin{figure}
\centering
\subfigure[\label{f:psdest1}]
{\includegraphics[width=2.5in,angle=0]{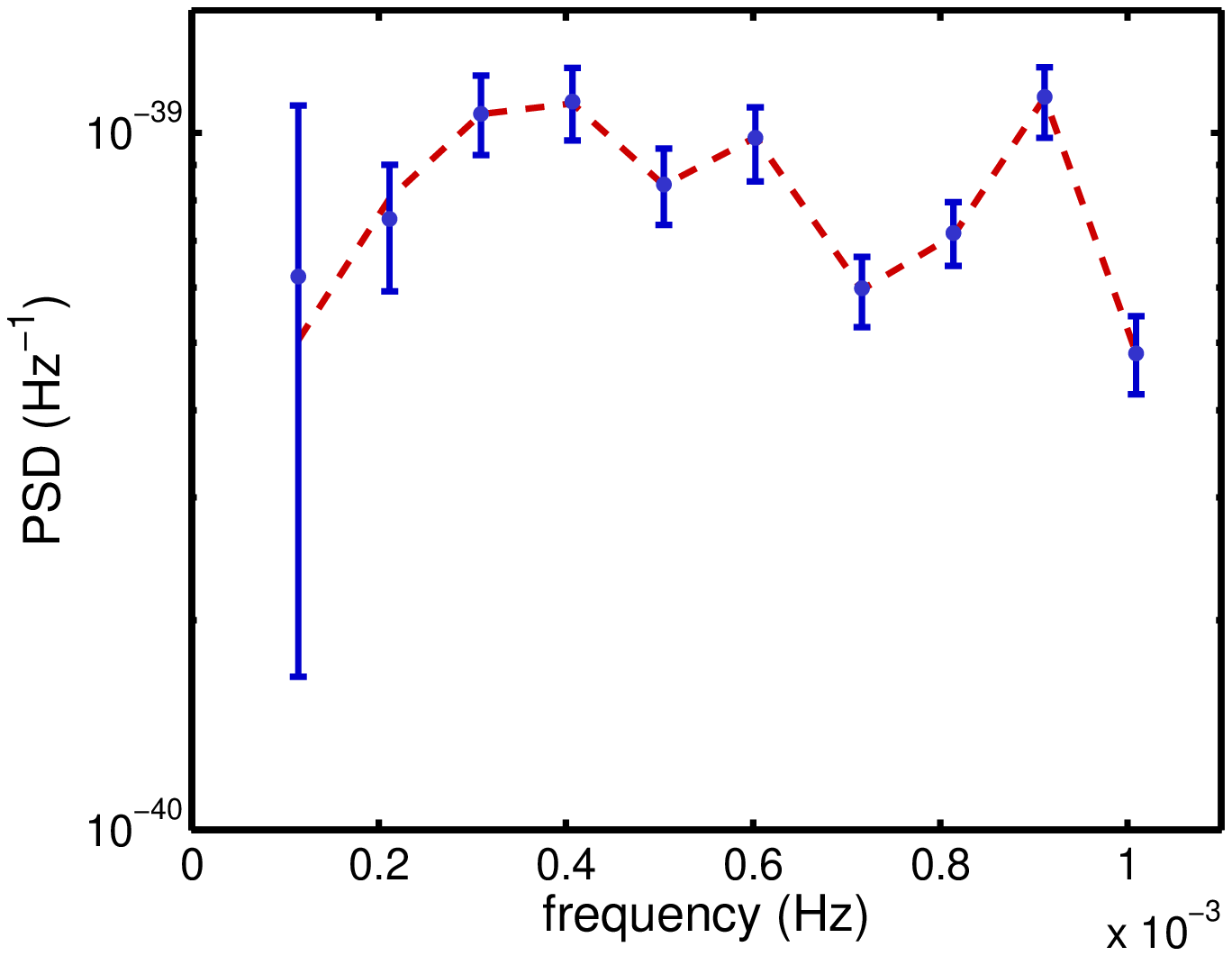}}
\subfigure[\label{f:psdest2}]
{\includegraphics[width=2.5in,angle=0]{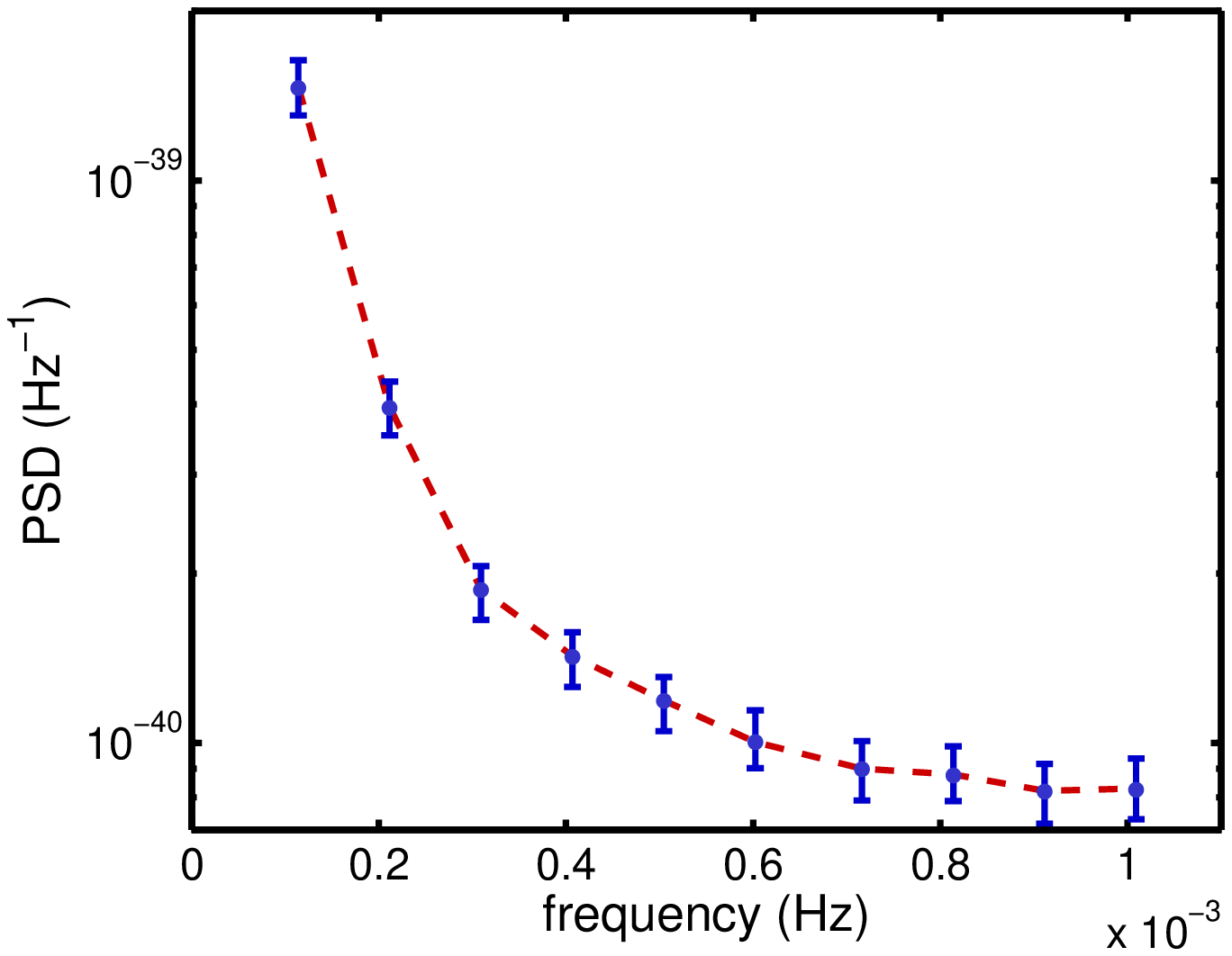}}
\caption{The estimated values of the PSDs of (a) the 
gravitational-wave signal, $P_h(f)$, and 
(b) the instrumental noise, $P_T(f)$, for ten different frequency bins. 
The points and error bars show the posterior means and 95\% probability 
intervals obtained from the posterior PDFs (scaled appropriately
from $\sigma_h^2$ and $\sigma_T^2$ respectively). The red dashed lines show 
the expected values of these PSDs as calculated from the 2.1 and 1.1.1.a
data sets.
}
\end{figure}

These priors were then used to calculate marginalised posterior PDFs 
for $\sigma_h^2$ and $\sigma_T^2$, from which the posterior mean and 
95\% probability interval in each band could be calculated.
These values were then scaled by a factor of $2\Delta t^2/T_\mathrm{seg}$ to
obtain (dimensionfull) estimates of the PSDs of the 
gravitational-wave signal $P_h(f)$ and the instrumental noise $P_T(f)$. 
Figure~\ref{f:psdest1} shows the Bayesian 95\% probability intervals 
for $P_h(f)$, and Fig.~\ref{f:psdest2} shows the corresponding
95\% probability intervals for $P_T(f)$.
From these figures, one sees that the Bayesian estimates agree with 
the expected PSD estimates to within the $95\%$ probability intervals.
In this band the strength of the stochastic signal 
$\sigma_h/\sigma_A$ varied from about 1 to $\approx 62$.  

%%%%%%%%%%%%%%%%%%%%%%%%%%%%%%%%%%%%%%%%%%%%%%%
\section{Conclusion}
\label{s:conclusion}

The observation of stochastic gravitational wave radiation 
from the early-Universe and/or astrophysical 
populations of sources at medium-to-high redshift may 
provide access to a large
portion of the discovery parameter space of LISA. 
Because this signal is isotropic, 
and the standard cross-correlation technique adopted in 
ground-based observations~\cite{AllenRomano:1999, LIGO-S4-stoch}
is insensitive to such radiation due to the geometry of the LISA 
pseudo-Michelson observables, a new analysis approach 
needs to be considered. 
The method that has been proposed so far relies on a 
particular combination of 
the LISA output, called the 
\emph{symmetrised Sagnac} channel~\cite{TintoEtAl:2000,HoganBender:2001}, 
which acts as a real-time noise monitor for LISA. 
To date, this approach has been put forward only at the conceptual level: 
it needs to be characterised quantitatively and then
further developed to produce an end-to-end analysis algorithm and 
pipeline that can enable the science exploitation of the LISA data.

The purpose of our study was to start to 
assess, quantitatively, the performance of the symmetrised Sagnac method;
in particular, we focused on the dependence of the method on
different levels of prior information regarding the instrumental 
noise,
and have presented here the first preliminary results of this study.
We developed a general approach in the framework of Bayesian 
inference, and an end-to-end analysis algorithm based on Markov 
Chain Monte Carlo methods to compute the posterior probability
density functions of the relevant model parameters. 
As a concrete example, we applied the method to data released 
as part of the second round of the Mock LISA Data 
Challenges~\cite{mldc, mldclisasymp, mldcgwdaw2, mldcamaldi2}. 

We showed that with no prior information about the ratio 
between the 
power of the instrumental noise contributions in the Sagnac 
and gravitational-wave-sensitive channels, it is indeed 
impossible to determine the presence of a stochastic 
gravitational-wave signal. 
However, for the specific cases considered here, 
moderate information (i.e., to within a factor $\approx 2$)
about the above ratio allows one not only to determine the 
existence of a signal, but also to place limits on 
the parameters that 
describe it. Indeed, it is likely that the LISA instrument will 
be very well
characterised, enabling searches for isotropic 
stochastic gravitational waves. 

In our view the most significant outcome of our study is a framework 
for more realistic investigations of LISA's performance and the
development of analysis techniques in the context of searches for 
stochastic signals. 
Our approach can be immediately extended in various ways;
first, by including the other independent and noise 
orthogonal TDI channel $E$ to improve the sensitivity of the
search, and second, by parameterising the spectra in order 
to study the frequency dependence of the gravitational-wave 
signal and instrumental noises. 
In addition, one can investigate how the prior 
knowledge of the instrumental noise levels required 
to unambigously detect a stochastic signal depends on the 
signal-to-noise ratio in the gravitational-wave-sensitive channel.
These extensions are currently under investigation and will be 
discussed in detail in a forthcoming paper \cite{RobinsonEtAl:2008}. 

%%%%%%%%%%%%%%%%%%%%%%%%%%%%%%%%%%%%%%%%%%%%%%%%%%%%%%%%%%%%%%%%%
\ack{The authors would like to thank C Ungarelli for  many 
fruitful conversations, from which this work originally started. 
ELR and AV 
acknowledge the support by the UK Science and Technology Facilities Council.
JDR acknowledges support of Leverhulme Trust Research Fellowship 
(2005/0104) and PPARC grant (PP/B500731) awarded to Cardiff University.}

%%%%%%%%%%%%%%%%%%%%%%%%%%%%%%%%%%%%%%%%%%%%%%%%%%%%%%%%%
\section*{References}

%bibliography


\begin{thebibliography}{99}

\bibitem{lisa}
Bender B L {\em et al.} 1998 
{\it LISA Pre-Phase A Report; Second Edition} MPQ 233

\bibitem{Hills-et-al:1990}
Hils D, ~Bender P L and Webbink R F 1990
{\em Astrophys.\ J.} {\bf 360} 75

\bibitem{BenderHills:1997}
Bender P L and Hils D 1997
{\em Class.\ Quantum Grav.} {\bf 14} 1439

\bibitem{HillsBender:2000}
Hils D and Bender P L 2000
{\em Astrophys.\ J.} {\bf 537} 334

\bibitem{Nelemans-et-al:2001}
Nelemans G, Yungelson L R and Portegies-Zwart S F 2001
{\em Astron. Astrophys.} {\bf 375} 890

\bibitem{Edlund-et-al:2005}
Edlund J A,  Tinto M, Kr{\'o}lak A and Nelemans G 2005
{\em Phys.\ Rev.\  D} {\bf 71} 122003

\bibitem{Maggiore:2000}
  Maggiore M 2000
  %``Gravitational wave experiments and early universe cosmology,''
  {\em Phys.\ Rept.}  {\bf 331} 283 
  %[arXiv:gr-qc/9909001].
  %%CITATION = PRPLC,331,283;%%

\bibitem{Hogan:2006}
  Hogan C J 2006
  %``Gravitational wave sources from new physics,''
  {\em AIP Conf.\ Proc.}\  {\bf 873} 30
  %[arXiv:astro-ph/0608567].
  %%CITATION = APCPC,873,30;%%
  
%\cite{Damour:2004kw}
\bibitem{DamourVilenkin:2005}
  Damour T and Vilenkin A 2005
  %``Gravitational radiation from cosmic (super)strings: Bursts, stochastic
  %background, and observational windows,''
  {\em Phys.\ Rev.\  D} {\bf 71} 063510 
  %[arXiv:hep-th/0410222].
  %%CITATION = PHRVA,D71,063510;%%

%\cite{Siemens:2006yp}
\bibitem{Siemens-et-al:2007}
  Siemens X, Mandic V and Creighton J 2007
  %``Gravitational wave stochastic background from cosmic (super)strings,''
  {\em Phys.\ Rev.\ Lett.}  {\bf 98} 111101
  %[arXiv:astro-ph/0610920].
  %%CITATION = PRLTA,98,111101;%%

\bibitem{FarmerPhinney:2003}
Farmer A J and Phinney E S 2003
{\em Mon.\ Not.\ R.\ Astron.\ Soc.} {\bf 346} 1197

\bibitem{BarackCutler:2004-emriconf}
Barack L and Cutler C 2004 
{\em Phys.\ Rev.\ D} {\bf 70} 122002

\bibitem{UngarelliVecchio:2001}
  Ungarelli C and Vecchio A 2001
  %``Studying the anisotropy of the gravitational wave stochastic background
  %with LISA,''
  {\em Phys.\ Rev.\  D} {\bf 64} 121501
  %[arXiv:astro-ph/0106538].
  %%CITATION = PHRVA,D64,121501;%%

\bibitem{Cornish:2001}
  Cornish N J 2001
  %``Mapping the gravitational wave background,''
  {\em Class.\ Quantum Grav.} {\bf 18} 4277 
  %[arXiv:astro-ph/0105374].
  %%CITATION = CQGRD,18,4277;%%

\bibitem{Seto:2004}
  Seto N 2004
  %``Annual modulation of the galactic binary confusion noise background and
  %LISA data analysis,''
  {\em Phys.\ Rev.\  D} {\bf 69} 123005
  %[arXiv:gr-qc/0403014].
  %%CITATION = PHRVA,D69,123005;%%

\bibitem{SetoCooray:2004}
  Seto N and Cooray A 2004
  %``LISA measurement of gravitational wave background anisotropy:  Hexadecapole
  %moment via a correlation analysis,''
  {\em Phys.\ Rev.\  D} {\bf 70} 123005
  %[arXiv:astro-ph/0403259].
  %%CITATION = PHRVA,D70,123005;%%

\bibitem{KudohTaruya:2005}
  Kudoh H and Taruya A 2005
  %``Probing anisotropies of gravitational-wave backgrounds with a  space-based
  %interferometer: Geometric properties of antenna patterns  and their angular
  %power,''
  {\em Phys.\ Rev.\  D} {\bf 71} 024025
  %[arXiv:gr-qc/0411017].
  %%CITATION = PHRVA,D71,024025;%%

\bibitem{TaruyaKudoh:2005}
  Taruya A and Kudoh H 2005
  %``Probing anisotropies of gravitational-wave backgrounds with a  space-based
  %interferometer. II: Perturbative reconstruction of a  low-frequency skymap,''
  {\em Phys.\ Rev.\  D} {\bf 72} 104015 
  %[arXiv:gr-qc/0507114].
  %%CITATION = PHRVA,D72,104015;%%

\bibitem{TintoEtAl:2000}
Tinto M, Armstrong J W, and Estabrook F B 2000
{\em Phys.\  Rev.\ D} {\bf 63} 021101(R)

\bibitem{HoganBender:2001}
Hogan  C J and Bender P L 2001
{\em Phys.\  Rev.\ D} {\bf 64} 062002

\bibitem{mldc}
{\tt http://astrogravs.nasa.gov/docs/mldc/}

\bibitem{mldclisasymp} 
Arnaud K A {\em et al.} (the MLDC Task Force) 2006 
{\em Laser Interferometer Space Antenna: 6th International LISA Symp. 
(Greenbelt, MD, 19--23 Jun 2006)} 
ed Merkowitz S M and Livas J C (Melville, NY: AIP) 619; 
{\em ibid.} 625 
(long version with lisaXML description: Preprint gr-qc/0609106)

\bibitem{mldcgwdaw2} 
Arnaud K A {\em et al.} (the MLDC Task Force) 2007 
{\em Class. Quantum Grav.} {\bf 24} S551

\bibitem{mldcamaldi2} 
Babak S {\em et al.} 
(the MLDC Task Force and Challenge 2 Participants) 2007 
{\em Proceedings of the 7th Amaldi Conference on Gravitational Waves 
(8--14 July 2007, Sydney, Australia)} 
Preprint arXiv:0711.2667

\bibitem{RobinsonEtAl:2008}
Robinson E, Romano J and Vecchio A, in prep.

\bibitem{TDI}
  Tinto M and Dhurandhar S V 2005
  %``Time-delay interferometry,''
  {\em Living Rev.\ Rel.} {\bf 8} 4 
  %[arXiv:gr-qc/0409034].
  %%CITATION = 00222,8,4;%%

\bibitem{mldcgwdaw12}
Babak S {\em et al.} 
(the MLDC Task Force and Challenge 1B Participants), 
``The Mock LISA Data Challenges: from Challenge 1B to Challenge 3" 
submitted to {\em Class. Quantum Grav.}

\bibitem{mldcgwdaw1} Arnaud K A et al. 
(the MLDC Task Force and Challenge 1 participants) 2007 
{\em Class. Quantum Grav.} {\bf 24} S529

\bibitem{Prince-et-al:2002}
  Prince T A, Tinto M, Larson S L and Armstrong J W 2002
  %``The LISA optimal sensitivity,''
  {\em Phys.\ Rev.\  D} {\bf 66} 122002
  %[arXiv:gr-qc/0209039].
  %%CITATION = PHRVA,D66,122002;%%

\bibitem{AllenRomano:1999}
Allen B and Romano J D  1999
%  %``Detecting a stochastic background of gravitational radiation: Signal
%  %processing strategies and sensitivities,''
  {\em Phys.\ Rev.\  D} {\bf 59} 102001
%  [arXiv:gr-qc/9710117].
%  %%CITATION = PHRVA,D59,102001;%%

\bibitem{LIGO-S4-stoch}
Abbott B \ et al. [The LIGO Scientific Collaboration] \ 2007  
{\em Astrophys.\ J.}  {\bf 659} 918

\end{thebibliography}
\end{document}